\documentclass{iaus}
\usepackage{graphicx}

\newcommand{\lsim}{\,\lower2truept\hbox{${<\atop\hbox{\raise4truept\hbox{$\sim$}}}$}\,} 
\newcommand{\gsim}{\,\lower2truept\hbox{${>\atop\hbox{\raise4truept\hbox{$\sim$}}}$}\,}

\title[Dark energy and CMB bispectrum]{Dark energy and CMB bispectrum}

\author[Fabio Giovi \& Carlo Baccigalupi]
{Fabio Giovi$^{1,2}$ and Carlo Baccigalupi$^{1,2}$}

\affiliation{$^1$SISSA/ISAS, Via Beirut 2-4, 34014 Trieste, Italy\\[\affilskip]
$^2$INFN, Via Valerio 2, 34127 Trieste, Italy\\
email: giovi@sissa.it; bacci@sissa.it}

\pubyear{2004}
\volume{225}
\pagerange{1--8}
\date{?? and in revised form ??}
\jname{Impact of Gravitational Lensing on Cosmology}
\editors{Mellier, Y. \& Meylan,G. eds.}
\begin{document}

\maketitle

\begin{abstract}
We consider the CMB bispectrum signal induced by structure formation
through the correlation between the Integrated Sachs-Wolfe and the
weak lensing effect. We investigate how the bispectrum knowledge can
improve our knowledge of the most important cosmological parameters,
focusing on the dark energy ones. The bispectrum signal arises at
intermediate redshifts, being null at present and infinity, and is
characterized by a large scale regime (dominated by linear dynamics of
cosmological perturbation) and a small scale one (dominated by density
perturbations in a non-linear regime); on the other hand, the effect
induced by dark energy on the power spectrum is mostly geometrical and
imprinted at redshift close to the present. Because of this, the
knowledge of power spectrum and bispectrum yield two complementary
informations at very different cosmological epochs, particularly
suitable to extract informations about the onset of the cosmic
acceleration and dark energy properties that provide it. In order to
quantify how  much the bispectrum can help the power spectrum in
constraining the dark energy parameters, we choose a fiducial model on
a three-dimensional space including the following dark energy
parameters: dark energy density $\Omega_V$; dark energy equation of
state today $w_0$ and dark energy equation of state in the past
$w_\infty$ ($w_\infty - w_0$ is related to the first derivative of
equation of state). Then we simulate a likelihood analysis showing how
contour levels become narrower when bispectrum is
included. Preliminary results suggest a consistent improvement on the
estimation of dark energy abundance and on dynamical properties of the
equation of state. This indicates that the knowledge of the bispectrum
in future high resolution and high sensitivity CMB observations could
yield a substantial improvement with respect to the traditional
analysis based on the power spectrum only. 
\end{abstract}

\firstsection % if your document starts with a section,

\section{Introduction}
The combination of several independent cosmological datasets, namely
type Ia supernovae (\cite{riess,perlm}), CMB (\cite{spergel}) and
large scale structure (see e.g. \cite{dodel}) indicate that the
universe is presently accelerating. The dark energy responsible for
the acceleration appears to be a fraction of about 73\% of the cosmic
energy density today. The main properties of the dark energy are
described in terms of its equation of state $w=p/\rho$, i.e. the ratio
between pressure and energy density; its value, for a pure
cosmological constant, is -1. A current experiments indicate that the
present value of $w$ should be in the range $w_0<-0.78$
(\cite{spergel}). The cosmological constant as an explanation to
cosmic acceleration has two well-known problems: the coincidence
problem (why cosmological constant density and matter density are
comparable today?) and the fine tuning problem (i.e. why cosmological
constant is 123 orders of magnitude less than Planck scale?). To solve
the latter problem, a dynamical scalar field, known as Quintessence
(\cite{peebl}), has been introduced as a minimal extension of the
cosmological constant; the dark energy equation of state gets
dynamical alleviating the fine tuning problem. In most models the dark
energy equation of state can be easily parameterized with only two
parameters: $w_{0}$ and its first derivative with respect to the scale
factor (\cite{linder}). The next challenge in cosmology is to
constrain the time evolution of $w$; this can be done with future SNIa
observations and future CMB experiments like Planck (\cite{balbi}). As
we shall see in the next section, the CMB power spectrum alone is
limited to constrain dark energy. Here we study the improvement which
might be achieved by taking into account the non-Gaussian distortion
induced by the correlation between weak lensing and Integrated
Sachs-Wolfe effect (ISW). Such signal is suitable studied in the
higher order statistic of CMB anisotropies. We choose here the CMB
bispectrum as an estimator of third order statistics (see
e. g. \cite{giovi}). The weak lensing effect on CMB anisotropies has
been studied (see e. g. \cite{komatsu} and references therein) and the
third order statistics, the bispectrum, has been used to constrain the
effective dark energy equation of state by \cite{verde}.

\section{Removing the distance degeneracy with CMB
  bispectrum}\label{sec:distances}

The main problem of the CMB power spectrum in studying the dark energy
is the degeneracy that affects the distance of last scattering surface
with respect the dark energy abundance and, in particular, its
equation of state. Its variation produces a change to the distance at
last scattering surface (see e. g. \cite{bacci}); unfortunately such
distance is degenerate with respect the main dark energy
parameters. This can be easily understood writing the formula of the
comoving distance to the last scattering surface (we restrict our
analysis to the flat case and we neglect the radiation contribution): 
\begin{equation}\label{e:dist}
r(z_{lss}) = \frac{c}{H_0} \int_0^{z_{lss}}\frac{dz}{\sqrt{\Omega_{0M}
    (1+z)^3 + \Omega_V e^{f(z)}}} .
\end{equation}
In the previous equation $c$ is the speed of light, $H_0$ is the
Hubble constant today, $z_{lss}$ is the redshift of last scattering
surface, $\Omega_{0M}$ is the matter density today,
$\Omega_V=1-\Omega_{0M}$ (because we consider only the flat case) is
the dark energy density and $f(z)$ depends on the equation of state of
dark energy $w(z)$ and is defined as
\begin{equation}\label{e:fz}
f(z) = 3 \int_0^{z'} dz'\frac{1+w(z')}{1+z'} .
\end{equation}
Analyzing eq. (\ref{e:dist}) and (\ref{e:fz}), we can see that the time
dependence on equation of state is washed out by two redshift integrations. 
In most models the dark energy equation of state can be parameterized with 
the following relation (\cite{linder})
\begin{equation}
w(z) = w_0 + (w_\infty - w_0) \frac{z}{1+z},
\end{equation}
where $w_0$ and $w_\infty$ are respectively its present and the asymptotic
values. The difference $(w_\infty - w_0)$ represents the time-variation of
dark energy equation of state. 

\begin{figure}
  \centering
  \resizebox{\hsize}{!}{\includegraphics[clip=true]{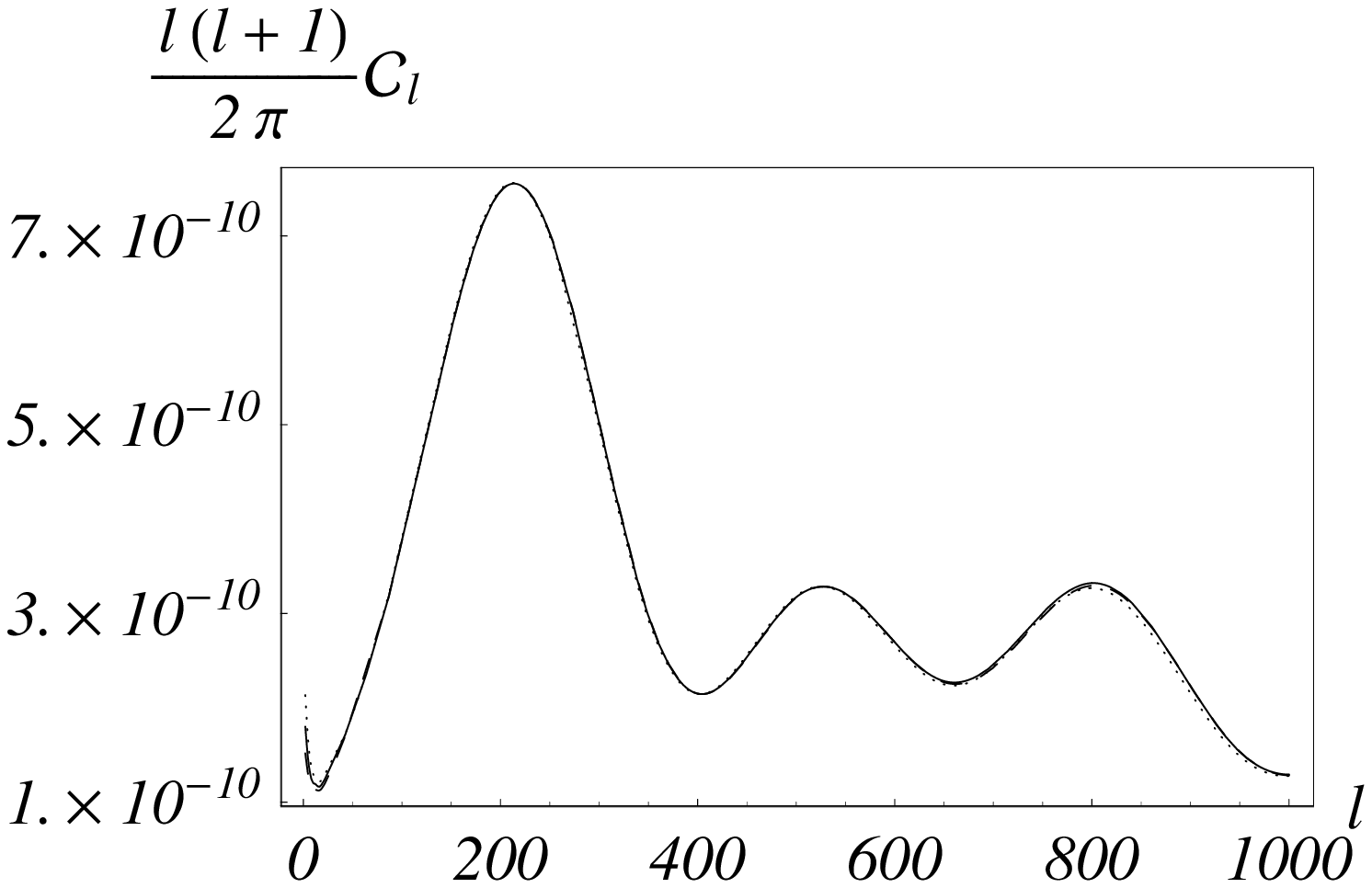}
    \includegraphics[clip=true]{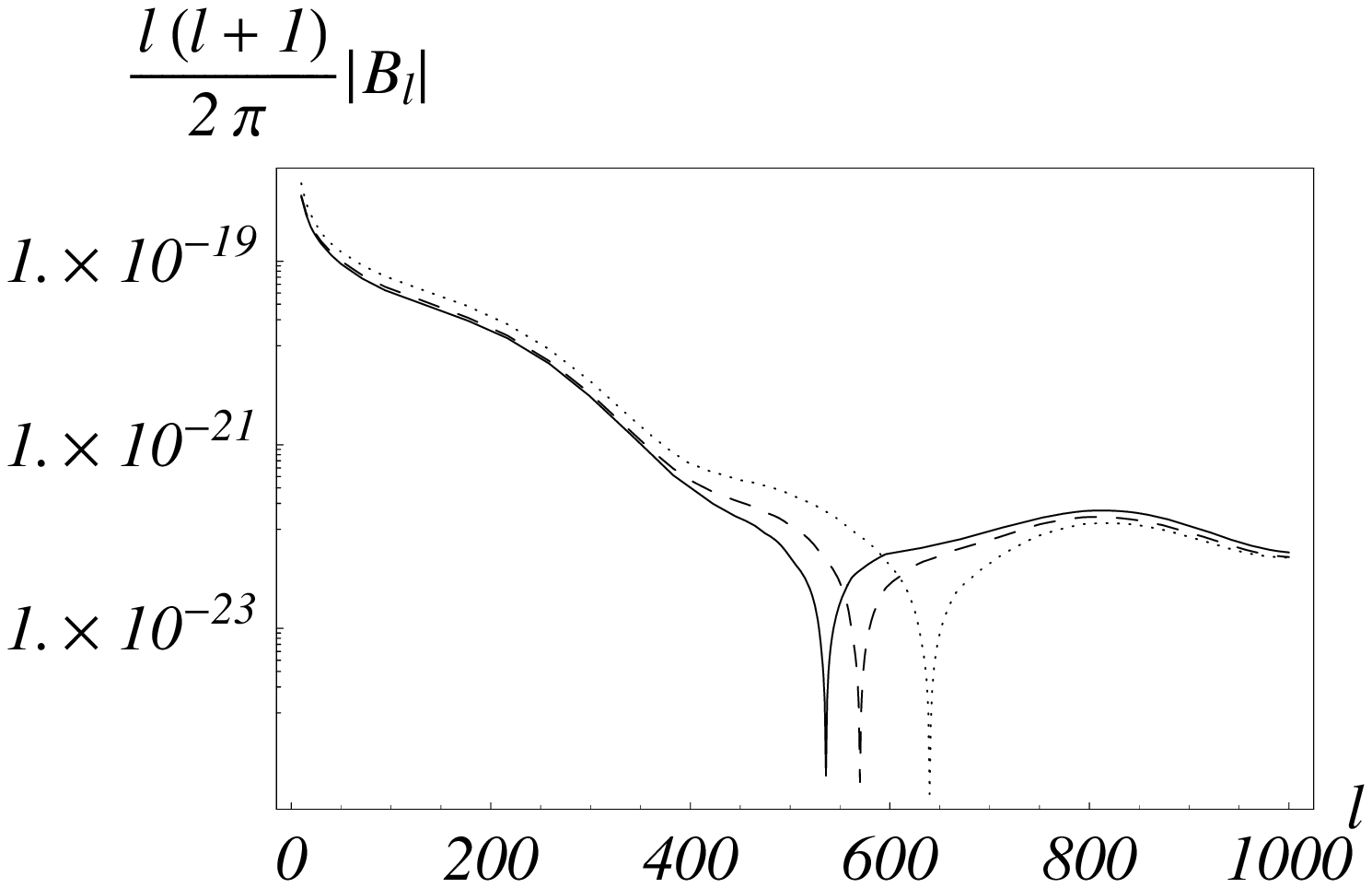}}
  \caption{CMB power spectra (left panel) and absolute value of
    equilateral bispectra (right panel) for models with same comoving
    distance of last scattering surface but different values of dark
    energy parameters. Solid line: $(\Omega_V=0.73, w_0=-1,
    w_\infty=-1)$. Dashed line: $(\Omega_V=0.735, w_0=-0.93,
    w_\infty=-0.89)$. Dotted line: $(\Omega_V=0.74, w_0=-1,
    w_\infty=-0.59)$. Notice that the three models are fully
    degenerate with the power spectrum while the degeneracy is
    removed with the equilateral bispectrum.}
  \label{f:cmb}
\end{figure}

Several different combinations of these dark energy parameters can produce the
same comoving distance to the last scattering surface; for example a comoving
distance of about 13900 Mpc can be obtained from these three sets of
dark energy parameters: $(\Omega_V=0.73, w_0=-1, w_\infty=-1)$;
$(\Omega_V=0.735, w_0=-0.93, w_\infty=-0.89)$ and $(\Omega_V=0.74,
w_0=-1, w_\infty=-0.59)$. The net effect of this distance degeneracy is
reflected in the CMB power spectrum; different cosmological models with the
same $r_{z_{lss}}$, will produce very similar power spectra (see
figure \ref{f:cmb}, right panel). Together with degeneracy in the
projection effect, there is another physical motivation for which the
CMB power spectrum is limited in its capability to constrain the dark
energy: the power spectrum is mostly injected at decoupling, and at
that time the dark energy density was negligible with respect to the
matter density and couldn't produce remarkable signatures in the CMB.

A way to include the CMB sensitivity on the dark energy is to consider the
signatures in the CMB produced by the correlation between the Integrated
Sachs-Wolfe effect and the weak lensing on the CMB. The ISW takes into
account the Rees-Sciama effect that arises when the non linear growth of
structures is included. The ISW effect affects the CMB photons with a
reddening; the photon acquires a blueshift when it falls down into the
potential well of growing structures and it acquires a redshift when it climbs
out, but these two contributions are not balanced because the
perturbations change in time. The gravitational lensing effect is the
well-known deflection of light due to the gravitational potential of
clump of matter. Both effects arise at the epoch of structure
formation, thus we can try to exploit them to alleviate the distance
degeneracy. The correlation between ISW and weak lensing induce a non
vanishing power in the CMB higher order statistics, which we study it
considering the CMB bispectrum. The latter is the harmonic transform
of three point correlation function and it is null, within cosmic
variance, if the CMB anisotropies are Gaussian. Since the ISW and the
weak lensing are produced by the same physical entity (growing
perturbations in the matter distribution), these secondary
anisotropies are correlated and induce a non-vanishing bispectrum with
exceeds the cosmic variance. 

\section{Integrated Sachs-Wolfe and weak lensing induced CMB
    bispectrum}\label{sec:bispectrum} 

When we consider the ISW and weak lensing effect, the CMB anisotropies
in a direction $\hat{n}$ in the sky can be decomposed as
\begin{equation}\label{e:dt}
\Theta(\hat{n}+\vec{\alpha}) \simeq
\Theta(\hat{n})+\vec{\nabla}\Theta\cdot\vec{\alpha}
\end{equation}
where $\Theta$ includes the primordial and ISW anisotropy contributions,
and the last term is the contribution from the weak lensing re-mapping;
$\vec{\alpha}$ is the deflection angle. Following \cite{verde} and
expanding eq. (\ref{e:dt}) in spherical harmonics, from the general
bispectrum definition $b_{l_1 l_2 l_3}^{m_1 m_2 m_3} = a_{l_1 m_1}
a_{l_2 m_2} a_{l_3 m_3}$, we can build the quantity
\begin{eqnarray} \label{e:genbis}
\nonumber
B_{l_1 l_2 l_ 3} & = & \sum_{m_{1}m_{2}m_{3}}\left(
\begin{array}{ccc}
l_1 & l_2 & l_3 \\
m_1 & m_2 & m_3
\end{array}
\right) b_{l_1 l_2 l_3}^{m_1 m_2 m_3} \\
& \simeq &
\sqrt{\frac{(2l_1+1) (2l_2+1) (2l_3+1)}{4\upi}} 
\left(
\begin{array}{ccc}
l_1 & l_2 & l_3 \\
0 & 0 & 0
\end{array}
\right) \cdot \\
\nonumber
& \cdot &\frac{l_1 (l_1 + 1) - l_2 (l_2 + 1) + l_3 (l_3 + 1)}{2}
C_{l_1}^{P} Q(l_3) + 5P.\ ,
\end{eqnarray}
where $5P.$ indicates the permutations over the three multipoles, the
parenthesis are the Wigner's 3J symbols, $C_l^P$ is the primordial CMB
power spectrum, and 
\begin{equation} \label{e:ql}
Q(l) \equiv <(a_{lm}^{lens})^* a_{lm}^{ISW}> \simeq 2 \int_0^{z_{lss}}
dz \frac{r(z_{lss}) - r(z)}{r(z_{lss})r^3 (z)} \left[ \frac{\partial
    P_{\Psi} (k,z)}{\partial z} \right]_{k=\frac{l}{r(z)}}\ , 
\end{equation}
is the correlation between lensing and ISW. In eq. (\ref{e:ql})
$P_{\Psi} (k,z)$ is the gravitational potential power spectrum; to
evaluate the non linear contribution to density power spectrum we have
used the existing semi-analytical approach (\cite{ma}). 

The asymptotic redshift behavior of the integrand of eq. (\ref{e:ql}), for $z
\rightarrow 0$ and $z \rightarrow \infty$, is vanishing (\cite{giovi}); in
fact, fixing the multipole, at low redshift the gravitational potential power
spectrum probes infinite wavenumbers where the power is vanishing, while at
high redshift the gravitational potential is constant since the universe
approach the standard CDM. Therefore, the bispectrum signal is acquired at
intermediate redshifts only, and is expected to reflect the cosmological
expansion rate at that epoch. The dark energy domination occurs approximately
at the same time, and therefore the bispectrum can be used as a tool to
investigate the dark energy properties and to alleviate the distance
degeneracy (\cite{giovi}). This can be seen clearly in figure \ref{f:cmb}
where we compare the CMB power spectrum with the equilateral bispectrum
$(l_1=l_2=l_3)$ for three dark energy models with the same comoving distance
at last scattering surface: the degeneracy is removed at the bispectrum level;
we describe in the next section the shape of bispectrum curves.

\section{Bispectrum features and likelihood analysis}\label{sec:like}

\begin{figure}
  \centering
  \resizebox{\hsize}{!}{\includegraphics[clip=true]{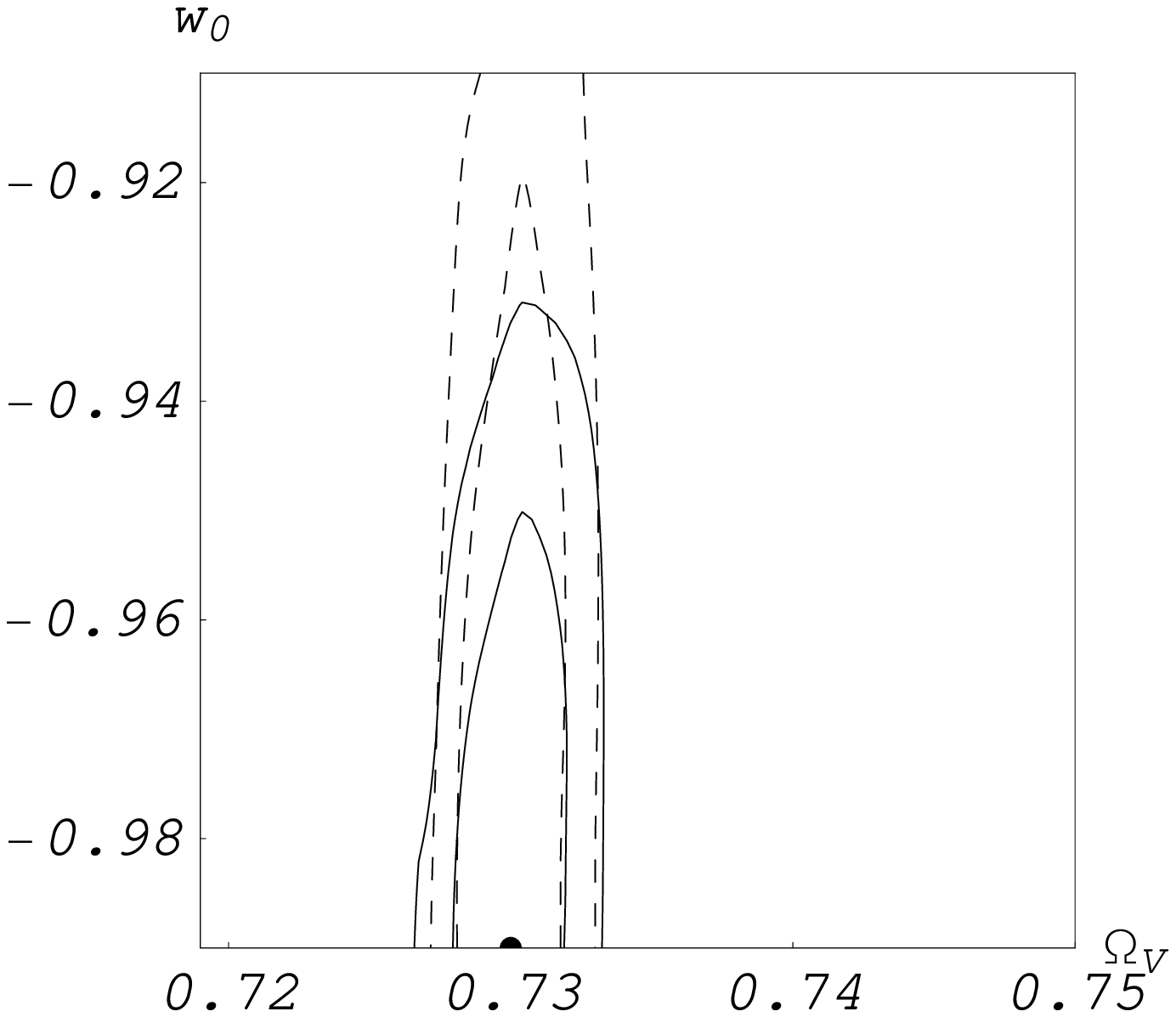}
    \includegraphics[clip=true]{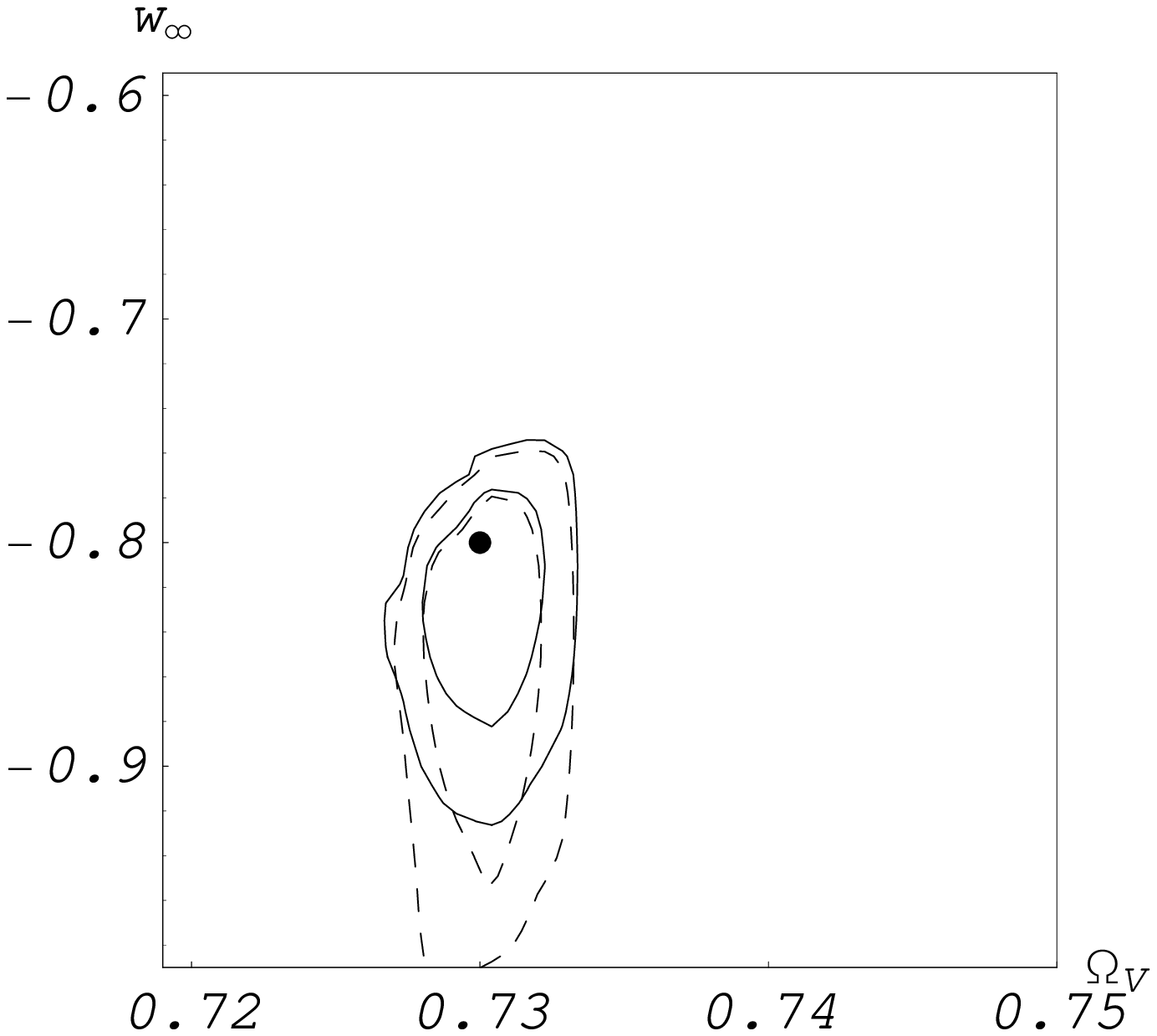}
    \includegraphics[clip=true]{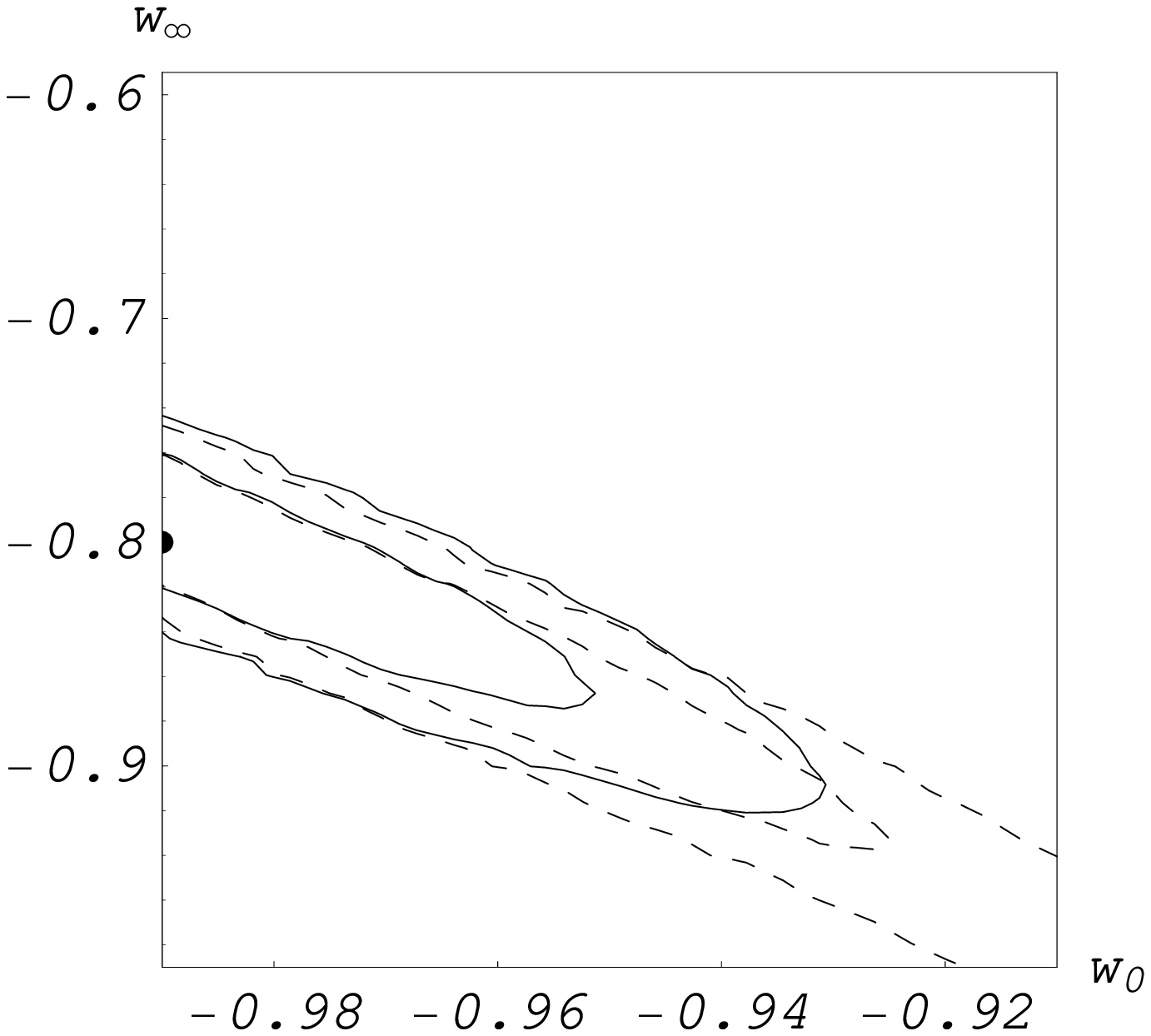}}
  \caption{Likelihoods confidence levels at 68\% (innermost contours)
    and 95\%(outermost contours) for power spectrum only (dashed line)
    and for power spectrum and bispectrum (solid line). From left to
    right: marginalization over $w_\infty$, marginalization over
    $w_0$, marginalization over $\Omega_V$. The filled dot is our
    fiducial model.}
  \label{f:like}
\end{figure}

The bispectrum has some peculiar features that can be used to
discriminate between different cosmological models. Plotting the
absolute value of bispectrum we have a \textit{cusp}, for which the
bispectrum is vanishing, and its position depends on cosmological
model (see right panel in figure \ref{f:cmb}). The position of this
zero-crossing point is the value of the multipole $l_0$ for which the
integrand of $Q(l)$ is null as shown \cite{verde}: the positive contribution
(due to linear growth) balances exactly the negative one (due to non linear
growth). Another relevant feature that must be taken into account is the
amplitude of the bispectrum in the linear part ($l<l_0$): different
cosmological models produce different amplitudes in the linear regime. The
linear power affect $l_0$: the higher is the power in the linear regime
($l<l_0$), the higher is the value of $l_0$.
\begin{table}
  \begin{center}
    \caption{Position and shift of multipole $l_0$ of bispectrum
    zero-crossing varying the main cosmological parameters.}
    \label{t:shift}
    \begin{tabular}{cccc}\hline
      Parameters & Lower & Higher & $\Delta l_0$ \\\hline
      $-1 \le w_0 \le -0.8$ & 418 & 482 & +66\\
      $0.60 \le \Omega_V \le 0.86$ & 324 & $>$1000 & $>$+676\\
      $0.020 \le \Omega_b  h^2 \le 0.028$ & 404 & 420 & +16\\
      $0.64 \le h \le 0.80$ & 492 & 360 & -132\\
      $0.80 \le n_s \le 1.12$ & 446 & 388 & -58\\\hline
  \end{tabular}
 \end{center}
\end{table}
In table \ref{t:shift} we analyze the variation of $l_0$ with respect to the
main cosmological parameters: dark energy equation of state today $w_0$, dark
energy density $\Omega_V$, baryons energy density $\Omega_b h^2$, Hubble
constant $h$ and spectral index of primordial fluctuations $n_s$. The values of
$l_0$ and its shift are evaluated when we vary each parameter, fixing
the others to a reference model. At first sight $l_0$ is more
sensitive to parameters which affect geometrically distances
$(\Omega_V, h)$, with the exception of $w_0$ at least in the interval
considered.
Unfortunately the bispectrum is much more noisy than the CMB power spectrum
already at the level of cosmic variance, since it is a higher order effect. We
can increase the signal to noise ratio including all triangles configurations
in $l$-space (i.e. considering all possible multipoles triplets) in our
analysis. We have tested the improvement that the bispectrum can bring to the
power spectrum alone building a three-dimensional likelihood on dark energy
parameters $(\Omega_V, w_0, w_\infty)$ and choosing a fiducial model for an
ideal (cosmic variance limited) experiment. In figure \ref{f:like} we show our
results marginalizing each time over one dark energy parameter; as we can see
in all cases the contours of the joint likelihoods are narrower than the
contours of the power spectrum likelihoods alone. This first, and very
preliminary result, is encouraging: a gain is expected on the estimation of
dark energy equation of state at the beginning of structure formation.

\section{Conclusions}\label{sec:conclusions}
We have discussed how to alleviate the distance degeneracy in the CMB power
spectrum using the ISW and weak lensing induced CMB bispectrum; we have shown
some preliminary results about the sensitivity of the bispectrum with
respect to the main cosmological parameters. We have simulated a joint
likelihood of power spectrum and bispectrum, limited to only dark energy
parameters and fixing a fiducial cosmological model. Our preliminary results
indicate that adding the bispectrum likelihood to the power spectrum one, the
contour levels are narrower and a gain in the estimation of dynamics
of dark energy equation of state is expected. Further study is needed to
assess the magnitude of this improvement.

Future precision cosmology data from high resolution experiments like Planck 
and future measures of tridimensional matter power spectrum with Ly-$\alpha$ 
forest (\cite{viel}) and cosmic shear (\cite{bacon}) will help in the use of
the bispectrum since they will allow on a better knowledge of the matter power
spectrum affecting the bispectrum signal.

%\pagebreak

\begin{acknowledgments}
Fabio Giovi is grateful to Francesco Valle for his hospitality in Lausanne
during the workshop.
\end{acknowledgments}


\begin{thebibliography}{}

\bibitem[{Baccigalupi \etal} 2002]{bacci}
  {Baccigalupi, C., Balbi, A., Materrese, S., Perrotta, F. \&
  Vittorio, N.} 2002, PRD, 65, 063520.

\bibitem[{Bacon \etal} 2004]{bacon}
  {Bacon, D. J. \etal}, preprint astro-ph/0403384.

\bibitem[{Balbi \etal} 2003]{balbi} 
  {Balbi, A., Baccigalupi, C., Perrotta, F., Materrese, S. \&
  Vittorio, N} 2003, ApJL, 588, L5.

\bibitem[{Dodelson \etal} 2002]{dodel} 
  {Dodelson, S. \etal} 2002, ApJ, {572}, 140.

\bibitem[{Giovi, Baccigalupi \& Perrotta} 2003]{giovi}
  {Giovi, F., Baccigalupi, C. \& Perrotta, F.} 2003, PRD, 68, 123002.  
  
\bibitem[Komatsu \& Spergel 2001]{komatsu} 
  {Komatsu, E. \& Spergel, D. N.} 2001, PRD, 63, 063002.

\bibitem[{Linder} 2003]{linder}
  {Linder, E.} 2003, PRL, 90, 091301.

\bibitem[{Ma \etal} 1999]{ma}
  {Ma, C. P., Caldwell, R. R., Bode, P. \& Wang, L.} 1999, ApJL, 521, L1.

\bibitem[{Peebles \& Ratra} 2003]{peebl} 
  {Peebles, P. J. E. \& Ratra, B.} 2003,
  \textit{Rev. Mod. Phys.} {75}, 599. 

\bibitem[{Perlmutter \etal} 1999]{perlm}
  {Perlmutter, S. \etal} 1999, ApJ, 517, 565.

\bibitem[{Riess \etal} 1998]{riess}
  {Riess, A. G. \etal} 1998, AJ, 116, 1009.
 
\bibitem[{Spergel \etal} 2003]{spergel}
  {Spergel, D. N. \etal} 2003, ApJS., 148, 175.

\bibitem[Verde \& Spergel (2002)]{verde} 
  {Verde, L. \& Spergel, D. N.} 2002, PRD, 65, 043007.

\bibitem[Viel, Heahnelt \& Springel 2004]{viel}
{Viel, M., Heahnelt, M. G. \& Springel, V.} preprint astro-ph/0404600.

\end{thebibliography}
\end{document}